\documentclass[10pt,conference]{IEEEtran}
\IEEEoverridecommandlockouts
\usepackage{cite}
\usepackage{amsmath,amssymb,amsfonts}
\usepackage{algorithmic}
\usepackage{graphicx}
\usepackage{comment}
\usepackage{textcomp}
\usepackage{xcolor}
\usepackage{url}
\def\BibTeX{{\rm B\kern-.05em{\sc i\kern-.025em b}\kern-.08em
    T\kern-.1667em\lower.7ex\hbox{E}\kern-.125emX}}

\usepackage[percent]{overpic} 

\usepackage[caption=false,font=footnotesize]{subfig} 

\begin{document}

\title{Securing Load Balancing over QUIC}




\author{
    \IEEEauthorblockN{
        Garegin Grigoryan\IEEEauthorrefmark{1}, 
        Dagim Mindaye\IEEEauthorrefmark{1},
        Shireen Maini \IEEEauthorrefmark{2},
        Minseok Kwon \IEEEauthorrefmark{2}
    }
    \IEEEauthorblockA{\IEEEauthorrefmark{1} Alfred University, Email: grigoryan@alfred.edu, dbm4@alfred.edu}
    \IEEEauthorblockA{\IEEEauthorrefmark{2} Rochester Institute of Technology, Email: sm4466@rit.edu, jmk@cs.rit.edu }
}

\maketitle
\begin{abstract}
In-network load balancing outperforms traditional software load balancing while costing less. For instance, programmable switch ASICs can use hashing to select the backend server for the initial packet of each flow at the line rate. However, when the pool of available servers changes, ensuring that the subsequent flow packets are mapped to the same server is challenging due to the data plane's limited memory resources and performance requirements. With the emergence of the QUIC transport protocol, several works show how Connection ID fields (CIDs) can embed the server identifier for all non-initial packets. This approach requires modifications on the server side and violates the QUIC specification, which mandates that CIDs remain unlinkable. In this work, we show that stateless QUIC load balancing can be implemented inside the data plane with no changes to CIDs. Moreover, QUIC packets, except the initial client packet, can bypass the load balancer. We also investigate and mitigate attacks on QUIC in this scenario, including full load balancer bypass and 0-RTT IP spoofing.
\end{abstract}

\section{Introduction}
\label{sec:intro}
Traffic load balancing across multiple backend servers helps achieve scalability and robustness of data center services. In a typical scenario, clients send their requests to a Virtual IP (VIP) assigned to the load balancer, which then selects the backend server. The load balancer should forward all the packets within a flow to the same server to maintain per-connection consistency (PCC). PCC can easily be achieved with a hash-based algorithm as long as the server pool remains unchanged. In realistic settings, however, changes in the pool or load-aware balancing strategies disrupt PCC.

In traditional software-based load balancing, a load-balancer middlebox stores the full mapping between the flow parameters (source and destination IP  addresses, transport layer protocol, and ports) and the selected server. With the emergence of programmable ASICs~\cite{bosshart2014p4}, significant efforts have been made to offload the load balancing function from the middleboxes to the data plane, offering reduced cost and lower latency~\cite{miao2017silkroad}. However, ensuring PCC in an in-network load balancer remains a challenge due to limited memory resources and the need to process packets at the line rate.

To illustrate this, consider a client-initiated TCP connection. It begins with a SYN packet sent to the in-network load balancer VIP and port. The load balancer then overwrites the IP and TCP header with the selected backend server address and forwards the packet to that server. If the server is not configured to use the VIP as its own source address, the load balancer has to rewrite the source IP in the server's SYN-ACK. Otherwise, a mismatch in IPs causes the client to drop the packet. Consequently, all subsequent client packets are destined to the VIP, and the switch must recover the correct backend server IP to preserve PCC. However, storing per-session mappings does not scale well given the limited in-switch memory. Moreover, the TCP header has no intrinsic field for the load balancer to embed server information that clients can echo in non-SYN packets. Attempts to address these challenges include offloading the flow-to-server mapping to match-action tables with the network controller assistance~\cite{miao2017silkroad}, or embedding a \textit{serverID} in the least significant bits of the TCP timestamp option~\cite{pit2018stateless}. The former deviates from a purely in-network load-balancing architecture, and the latter requires clients and servers to support the TCP timestamp option, and may reduce the accuracy of RTT-estimation mechanisms that rely on it.

\begin{figure}
	\begin{center}
		\includegraphics[width=0.9\columnwidth]{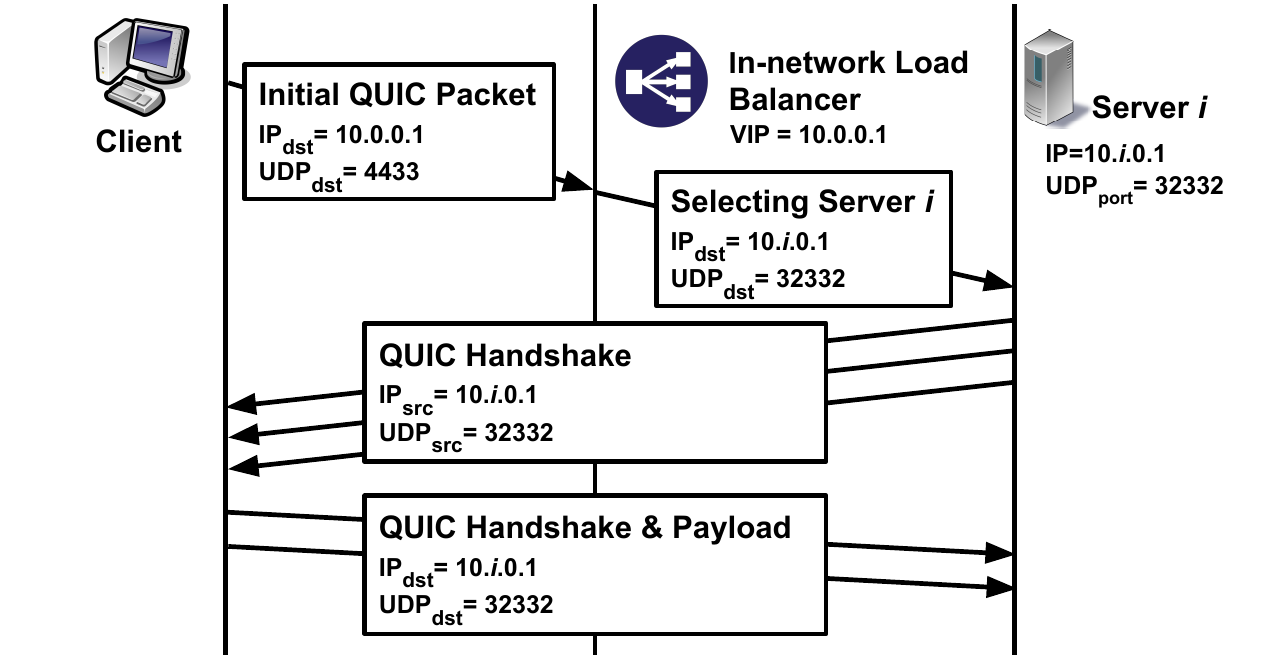}
		\caption{Load Balancing over QUIC protocol. The load balancer is involved only in handling the initial client packet to select a backend server}
        \label{fig:quiclb}
	\end{center}
        \vspace{-5mm}
\end{figure} 

In some prior work~\cite{langley2017quic}, QUIC is used for a stateless load balancer. Unlike TCP, QUIC has non-optional and unencrypted Connection ID (CID) fields (Source and Destination ID), which can be used to embed the \textit{serverID}. However, modifying CIDs requires changes on the QUIC server side, which might be difficult in multi-tenant environments where external users run their own deployments on a shared cluster. In addition, QUIC specifications mandate minimizing the linkability of QUIC packets to on-path observers~\cite{quicklinkability}.

In this paper, we present how to secure the QUIC load balancing in the programmable data plane. We leverage QUIC's support of changes in IP addresses and UDP ports without breaking the client-server session. While clients send initial packets to the Virtual IP and UDP port of the in-network load balancer and the servers reply with their own IP and port, no rewriting is performed by the load balancer. QUIC clients, following the QUIC specification, accept the replies and send their subsequent packets directly to the servers' IP addresses and ports (see Figure~\ref{fig:quiclb}). This approach requires no changes on the client or server's side and also minimizes the overhead of the load-balancing algorithm. Only the initial packet of each flow is processed by the load balancer, while subsequent packets are forwarded according to their destination IP address. 

However, this design introduces another challenge: how to prevent clients from bypassing the load balancer even for the initial packet, since the data plane must allow packets to be sent directly to server endpoints? Specifically, clients can learn the IP addresses and UDP ports of the backend servers and overload specific machines by directing their initial QUIC requests to them. Due to packet coalescing of QUIC headers and payloads, only deep packet parsing allows the stateless load balancer to identify whether a packet is the first packet of a new QUIC session. In this work, we show how to design such deep parsing in the programmable data plane. 

\begin{figure}
	\begin{center}
		\includegraphics[width=0.9\columnwidth]{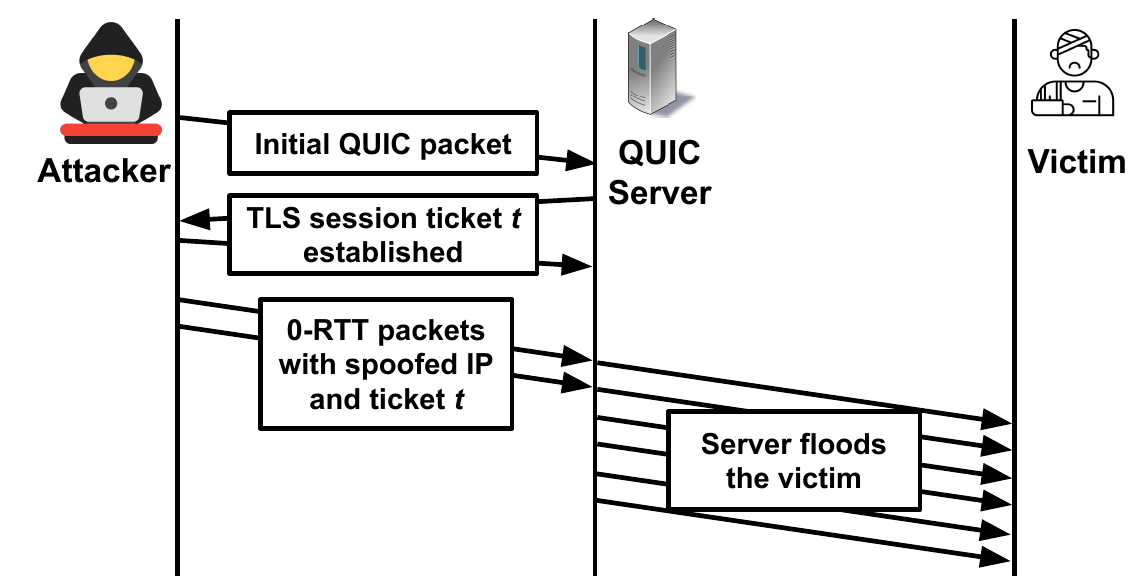}
		\caption{0-RTT attack flow }
        \label{fig:attack}
	\end{center}
        \vspace{-5mm}
\end{figure}

We also demonstrate how deep parsing can help mitigate 0-RTT attacks when a malicious user sends multiple packets with a spoofed source IP address (see Figure~\ref{fig:attack}). This attack exploits the flexibility of the QUIC protocol with dynamic IP addresses. Specifically, when a client request reuses the TLS session ticket established in the previous connection, the server begins sending the response without waiting for a second handshake to complete, even if the client's IP address and UDP port are different. When the server reply includes multiple data packets, all directed to the spoofed IP address, such attacks overwhelm both the server's and the victim's CPU and bandwidth. Techniques such as extra address validation or disabling 0-RTT require per-server changes and reduce performance~\cite{attack}. Our approach moves enforcement into the programmable network: deep parsing of QUIC packets allows detecting 0-RTT bursts, blocking them when they exceed a certain threshold. We also show that deep parsing has a negligible effect on the latency of QUIC flows. 

To summarize our contributions, we:
\begin{enumerate}
    \item Demonstrated QUIC's usability for load balancing with minimal load balancer involvement;
    \item Designed an in-network QUIC load balancer without control plane involvement and no server modifications;
    \item Leveraged deep QUIC parsing to mitigate attacks on QUIC clusters, such as full load balancer bypass and 0-RTT attacks with source IP spoofing;
    \item Evaluated the proposed system and demonstrated its effectiveness with no measurable increase in QUIC flows' latency.
\end{enumerate}

\section{Background}
\label{sec:bg}
QUIC~\cite{rfc9000} is built between the UDP header and the application layer (most commonly, HTTP/3). The unencrypted part of a QUIC header includes one bit that identifies the type of the header (long or short), two bits identifying the type of a packet (\textit{Initial}, \textit{Handshake}, \textit{0-RTT}, or \textit{Retry}), and fields of various byte-lengths identifying the version of QUIC protocol, the source and destination IDs, and the length of the payload.

QUIC session establishment combines transport and cryptographic handshakes (separate in TCP+TLS). It starts with the client's \textit{Initial} packet, followed by the server's \textit{Initial} and \textit{Handshake} packets. The client then sends another \textit{Initial} and \textit{Handshake} packets, after which the endpoints begin sending encrypted application data. To reduce latency on repeat connections, QUIC supports 0-RTT sessions: clients and servers can exchange application data if they have previously established a session, without waiting for the handshake to complete.

QUIC introduces a mechanism called packet coalescing, which allows multiple QUIC packets to be sent within a single UDP datagram. Coalescing is used during the handshake, allowing packets with types such as \textit{Initial}, \textit{Handshake}, and \textit{0-RTT} to be combined into a single datagram. This design reduces the overhead of multiple UDP/IP headers and minimizes the number of system calls required, resulting in lower latency during connection setup and higher network utilization. In the meantime, coalesced QUIC packets can be misclassified if classification relies only on the packet type field of the first QUIC header. Specifically, to classify the first \textit{Initial} QUIC packet (equivalent to SYN in TCP) for load balancing purposes, the parser needs to identify the \textit{Initial} packet that is not coalesced with the \textit{Handshake} packet. Furthermore, detecting 0-RTT traffic requires parsers to inspect the datagrams' bytes beyond the first coalesced QUIC packet.

\section{Design}
\label{sec:design}
\subsection{Load Balancing over QUIC}
\label{sec:lbquic}
We leverage the programmable data plane and Protocol-Independent Switch Architecture (PISA)~\cite{bosshart2014p4} to build an in-network load balancer for traffic over the QUIC transport protocol. In our implementation, programmability of layer 4 switches built on top of PISA allows the cluster operator to quickly adapt to changes in QUIC~\cite{muthuraj2024replication}. For implementation details, we utilize \textit{aioquic}, a popular open-source implementation of QUIC and HTTP/3~\cite{aioquic}, the P4Kube~\cite{grigoryan2025p4kube} cluster framework, and the PISA switch that load-balances traffic across Kubernetes deployments. 

Initially, the in-network load balancer is given the virtual IP addresses and ports used by clients to send connection-establishing QUIC packets, as shown in Figure~\ref{fig:quiclb}. Next, the load balancer selects a backend server from the available pool using ECMP-type hashing~\cite{cai2012rfc}, then overwrites the IP and UDP headers with the IP address and UDP port of the selected server.

\begin{figure}
	\begin{center}
		\includegraphics[width=0.9\columnwidth]{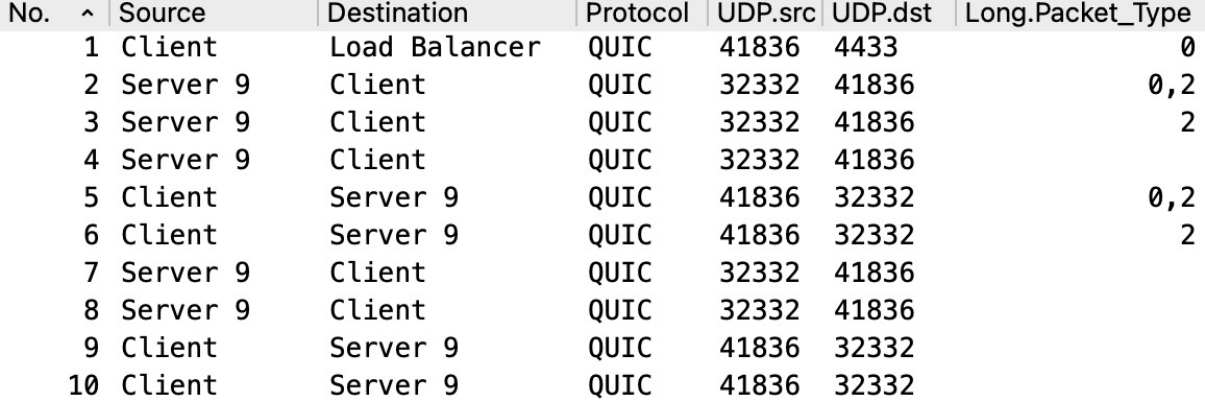}
		\caption{\label{fig:http3}QUIC+HTTP/3 request in P4Kube}
	\end{center}
    \vspace{-5mm}
\end{figure}

We analyzed the \textit{aioquic}'s QUIC+HTTP/3 trace, and interestingly found that a QUIC client accepts responses from servers without the intervention of the load balancer. An example is shown in Figure~\ref{fig:http3}. In the example, the client sends the initial packet to the VIP of the load balancer and UDP port $=$ 4433 (No. 1). The response from the backend server (in No. 2) arrives at the client's host, with the packet's source IP equal to server's original IP address 19.0.0.2, and source UDP port $=$ 32332, the server's UDP port. The subsequent client packets (see packets 5, 6, 9, and 10 in the trace) are addressed directly to the server, bypassing the load balancer and preserving PCC.

This is different from a typical TCP workflow where the client socket is bound to the load balancer's VIP address and TCP port, and the packets originating from a different destination address or port are rejected. To avoid this mismatch, the load balancer should overwrite the source IP address and TCP port for all the server packets. This inevitably adds latency for all egress packets. Worse, maintaining the correct backend mapping for all subsequent (i.e., non-initial) TCP packets from a client adds additional strain on switch resources~\cite{barbette2020high}. This is because these packets must be forwarded to the server selected earlier by the load balancer in accordance with the PCC requirement, even though the client continues using the Virtual IP and TCP port as the destination.

QUIC itself supports bypassing the load balancer after the initial client packet. This means that all the subsequent server and client packets can use the real IP address and UDP port of the backend server selected by the load balancer (as shown in Figures~\ref{fig:quiclb} and~\ref{fig:http3}). This workflow, however, introduces a potentially serious flaw: clients can target the chosen servers with their initial packets, fully bypassing the load balancer since the servers’ real IP addresses become known. 

\subsection{Preventing Load Balancer Bypass}
\begin{figure}

	\begin{center}

	\includegraphics[width=0.9\columnwidth]{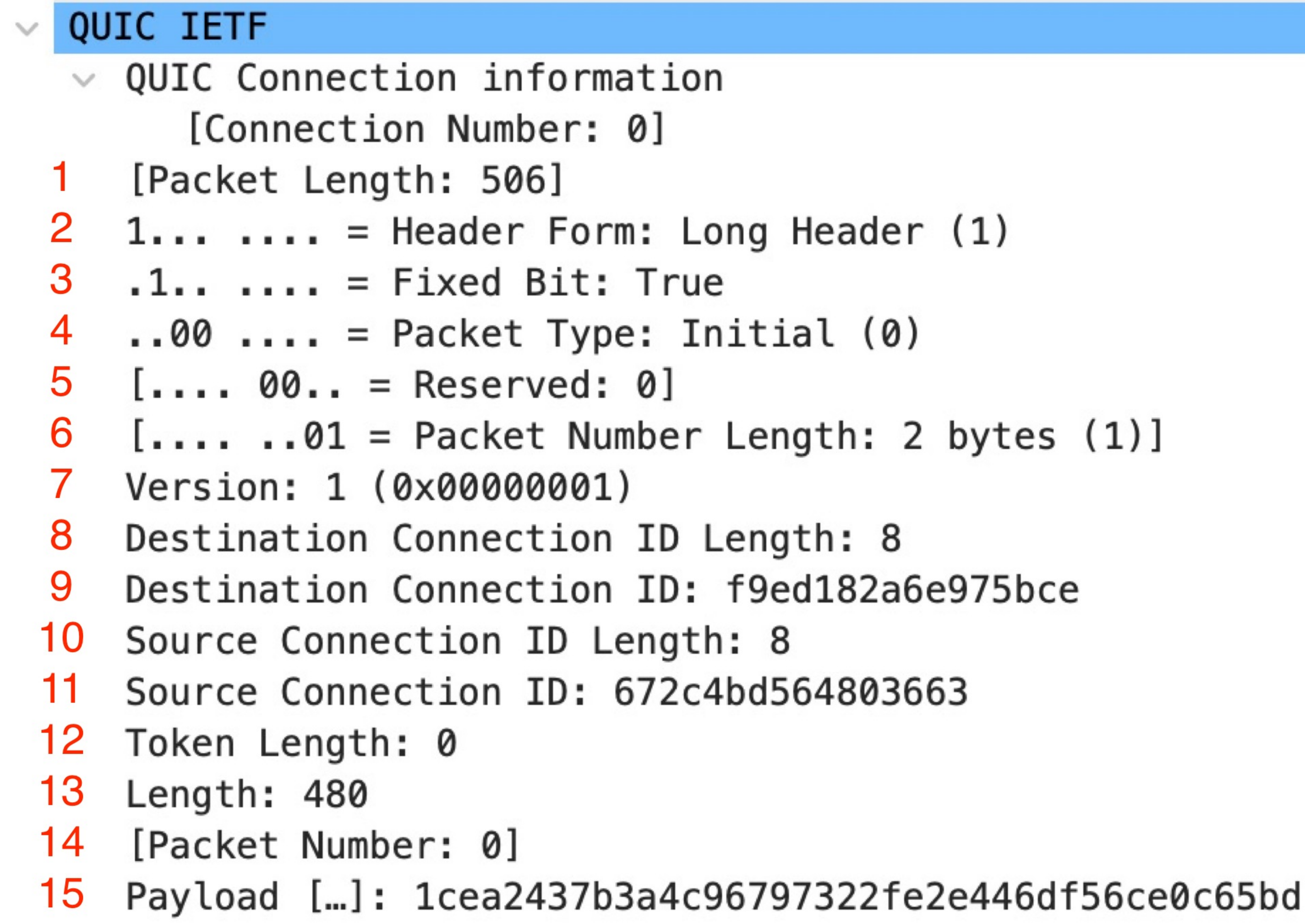}
		\caption{\label{fig:pkt1}QUIC header of the first client packet}
    
	    \vspace{-5mm}
	\end{center}
\end{figure} 

\begin{figure}
	\begin{center}
		\includegraphics[width=0.85\columnwidth]{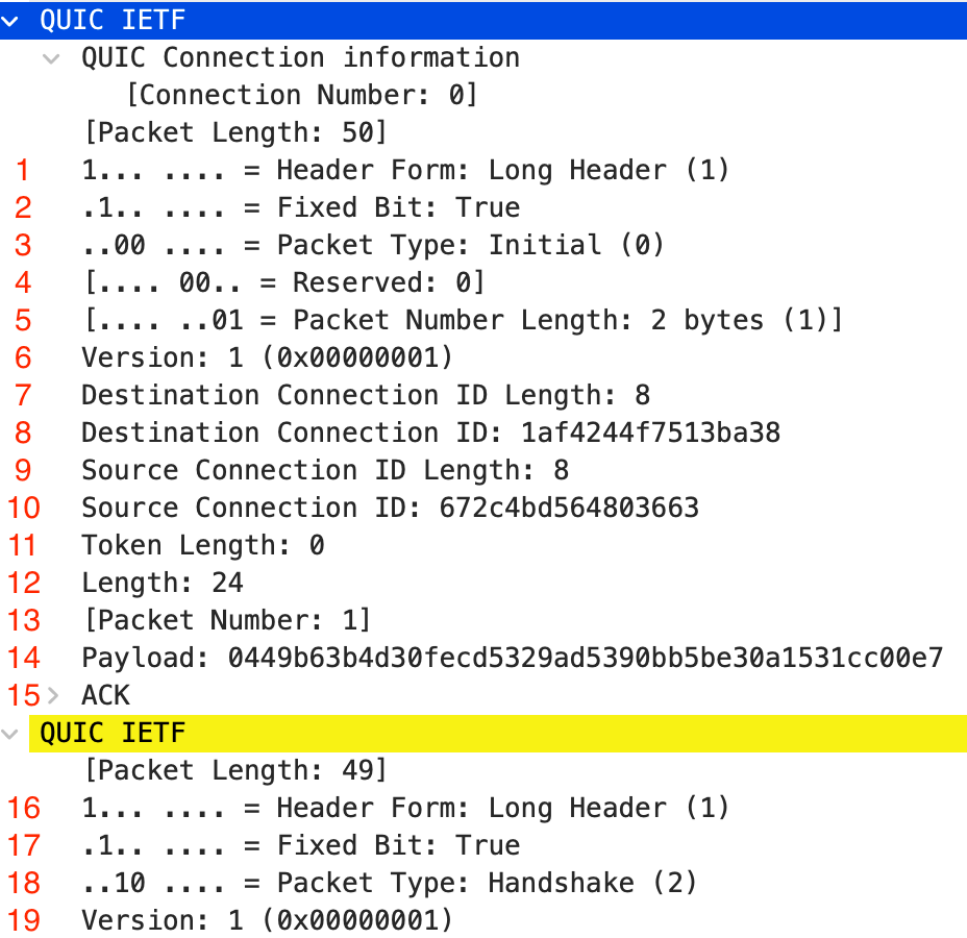}
		\caption{\label{fig:pkt5}QUIC headers of the second client packet}
	\end{center}
        \vspace{-5mm}
\end{figure} 
To prevent load balancer bypassing by connection-establishing packets, we need to detect and drop the first packets of each new connection if they are addressed directly to backend servers. In TCP, this is straightforward since the initial packets are explicitly marked by the SYN flag. In QUIC, however, the analogous \textit{Packet Type} flag is set to 0 (i.e., \textit{Initial}) for multiple handshake datagrams of QUIC clients. For instance, the fifth packet seen in the trace in Figure~\ref{fig:http3} would be dropped. To resolve this issue, we need to identify the differences in the QUIC long headers of the first and second client packets.

Figure~\ref{fig:pkt1} shows a Wireshark snapshot of the first client QUIC packet. It contains a single long QUIC header with the 2-bit field \textit{Packet Type} set to 0 (i.e., \textit{Initial}, see line 4). The next client packet from the trace (see Figure~\ref{fig:pkt5}) also contains a long QUIC header, with its \textit{Packet Type} field set to 0 like in the first packet (see line 3). The key difference is that this second packet contains a coalesced second long QUIC header with \textit{Packet Type} set to 2 (i.e., \textit{Handshake}, see line 18). This means that to prevent the load balancer bypassing by connection-establishing client packets (i.e., when the destination IP address of such packets is not the VIP of the load balancer), the PISA switch must verify the existence of the second long QUIC header.
\begin{figure}
	\begin{center}
		\includegraphics[width=0.7\columnwidth]{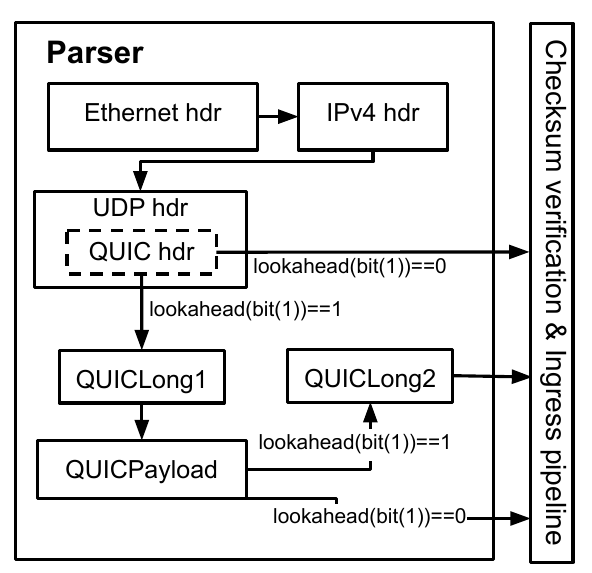}
		\caption{\label{fig:parser}Parser of PISA switch}
	\end{center}
        \vspace{-5mm}
\end{figure} 

 Unfortunately, simply comparing the UDP and QUIC payload length fields can lead to incorrect results, since QUIC specification~\cite{rfc9000} requires \textit{Initial} packets to be padded with null bytes to ensure a minimum UDP datagram size of 1200 bytes. In our design, we rely on PISA switches' look-ahead functionality to determine the existence of coalesced long QUIC packets (see Figure~\ref{fig:parser}). Specifically, by the time the parser advances to the UDP payload field, it looks ahead by one bit to determine whether the payload contains a long or a short QUIC header. To check the existence of the second QUIC longer header, the parser advances by the number of bits corresponding to the variable-length payload, and then again performs a look-ahead at the next bit. If that bit is also set to one, then the parser will be able to identify the second long QUIC header and its fields.

\begin{figure}
	\begin{center}
		\includegraphics[width=0.9\columnwidth]{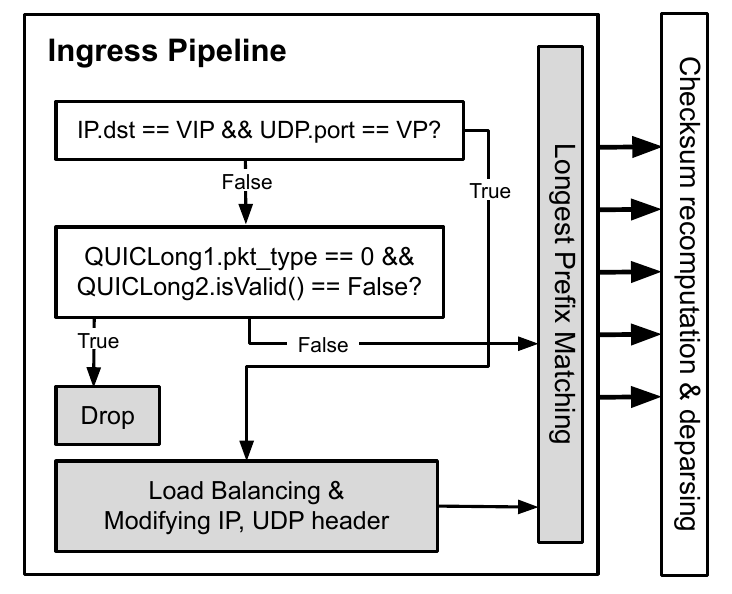}
		\caption{\label{fig:lbingress}Ingress pipeline for load balancing of QUIC traffic}
	\end{center}
        \vspace{-6mm}
\end{figure} 

Figure~\ref{fig:lbingress} shows the ingress pipeline of the QUIC load balancer. All the packets destined to the Virtual IP address and port of the load balancer are processed by the load-balancing algorithm (e.g., ECMP), with the packets' IP and UDP headers modified based on the selected server. For all the other QUIC packets, the switch checks whether it is an \textit{Initial} packet (i.e., packet type is 0) and does not contain the second header. In case both conditions are true, the packet is determined to be the first packet of a new flow, yet directed to a specific backend server. Such packets are dropped to prevent full bypassing of the load balancer. Otherwise, if the QUIC packet is not a flow-initiating packet, it is moved to the next stages (i.e., Longest Prefix Matching, checksum calculation, and deparsing and forwarding).

\subsection{0-RTT Attack Mitigation}
\begin{figure}
	\begin{center}
		\includegraphics[width=0.9\columnwidth]{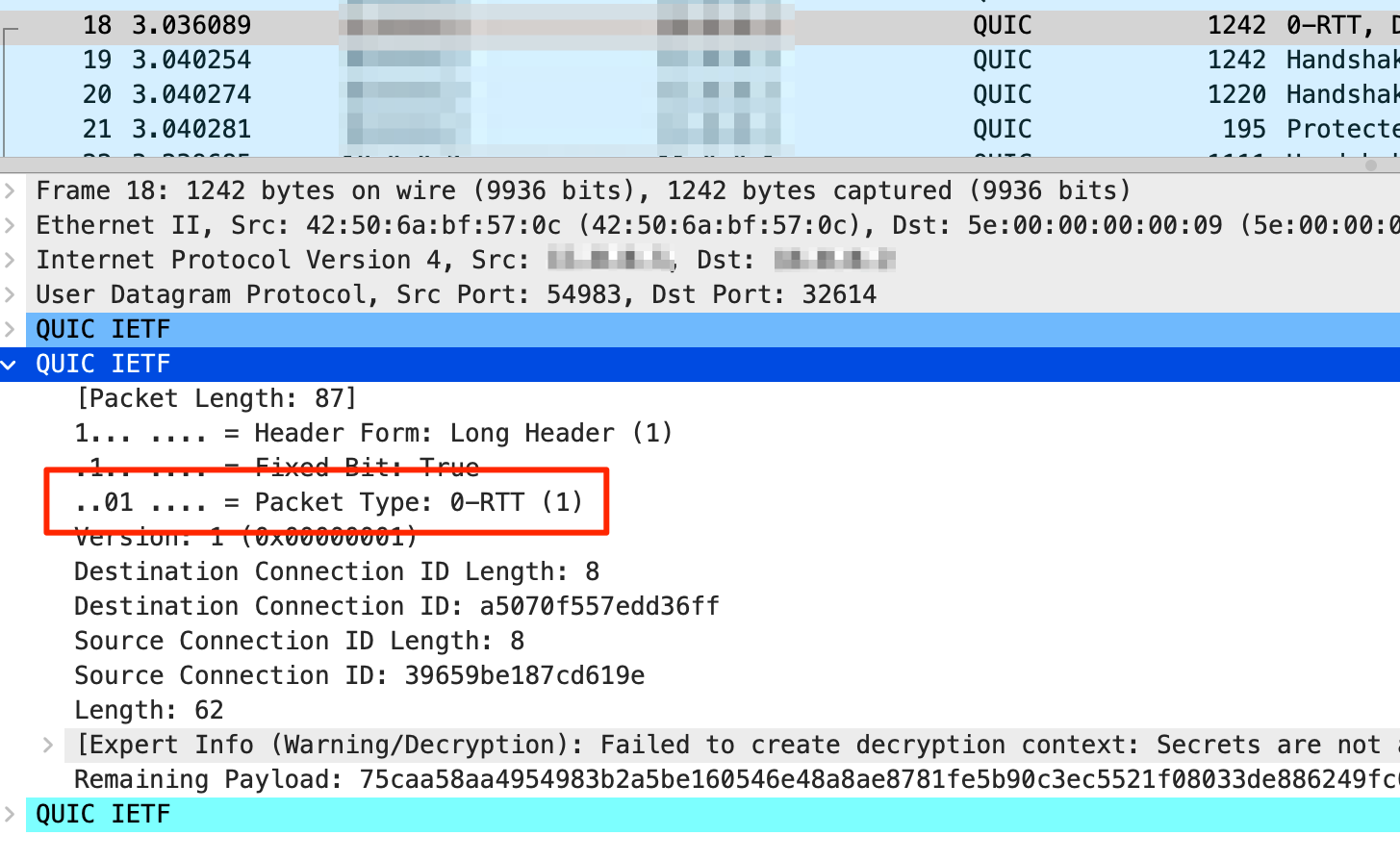}
		\caption{\label{fig:ws}Wireshark trace of a 0-RTT attack}
	\end{center}
    \vspace{-5mm}
\end{figure}
\begin{figure}
	\begin{center}
		\includegraphics[width=0.9\columnwidth]{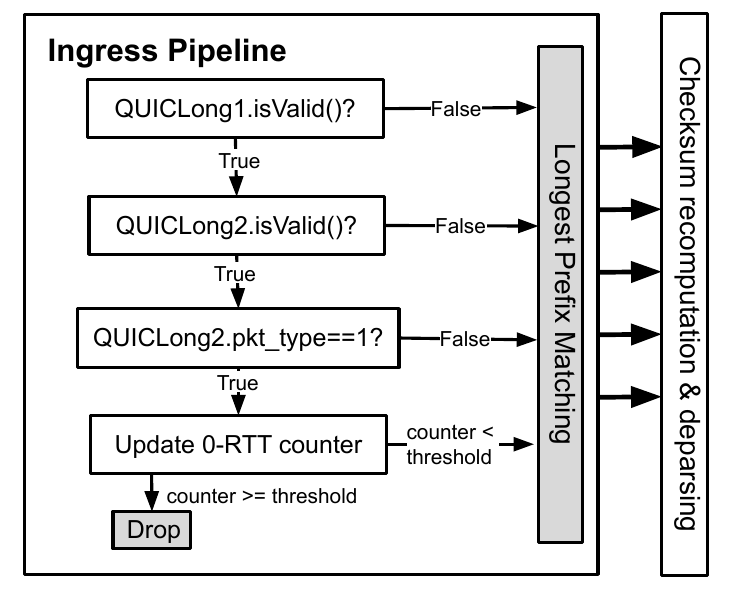}
		\caption{\label{fig:0rttingress}Ingress pipeline for 0-RTT attack mitigation}
	\end{center}
        \vspace{-6mm}
\end{figure}

 
A 0-RTT attack exploits QUIC's flexibility in allowing source IP address changes when the client reuses the cryptographic token established during the initial handshake. For later sessions, the client uses a victim IP address to send a request expecting amplified responses directed to the victim (see Figure~\ref{fig:attack}). Our analysis of a 0-RTT attack trace (see Figure~\ref{fig:ws}) shows that the programmable switch needs access to the second coalesced header of QUIC to identify 0-RTT packets, similar to the approach we used to prevent full load balancer bypass. In the ingress pipeline of the parser (see Figure~\ref{fig:0rttingress}), we count the 0-RTT packets and check the counter against a pre-defined threshold. If the counter exceeds the threshold, the 0-RTT packet is dropped, preventing the potential attack; otherwise, the packet is forwarded as normal. 
The counter is reset to zero after a certain amount of time (i.e., an epoch) to allow counting the number of 0-RTT bursts rather than the total non-decreasing number of 0-RTT occurrences. We measure the time by extracting the packets' ingress timestamp metadata and comparing it against the most recent epoch timestamp. This design allows line-rate detection of 0-RTT packets and stops 0-RTT attacks across the entire network behind the load balancer, rather than only at an individual server.

We note that determining the exact threshold value and the epoch length is outside the scope of this work.
\section{Evaluation}
\label{sec:eval}
\subsection{Experimental Setup}
We implemented the prototype load balancer and 0-RTT attack mitigation system in the P4 language for data planes with PISA architecture. Our experimental setup included:
\begin{itemize}
    \item \textit{FABRIC} testbed~\cite{fabric-2019}: a large-scale, programmable infrastructure to host the cluster. Our testbed slice contained 10 replica servers for load balancing QUIC requests. 
    \item \textit{aioquic}~\cite{aioquic}: a Python-based implementation of QUIC and HTTP/3 protocols.
    \item \textit{P4Kube}~\cite{grigoryan2025p4kube}: In-Network Load Balancer Framework for Kubernetes with \textit{bmv2} P4 switch emulator~\cite{bmv2}.
\end{itemize}

\subsection{Load Balancing over QUIC}
\begin{figure}
	\begin{center}
		\includegraphics[width=0.8\columnwidth]{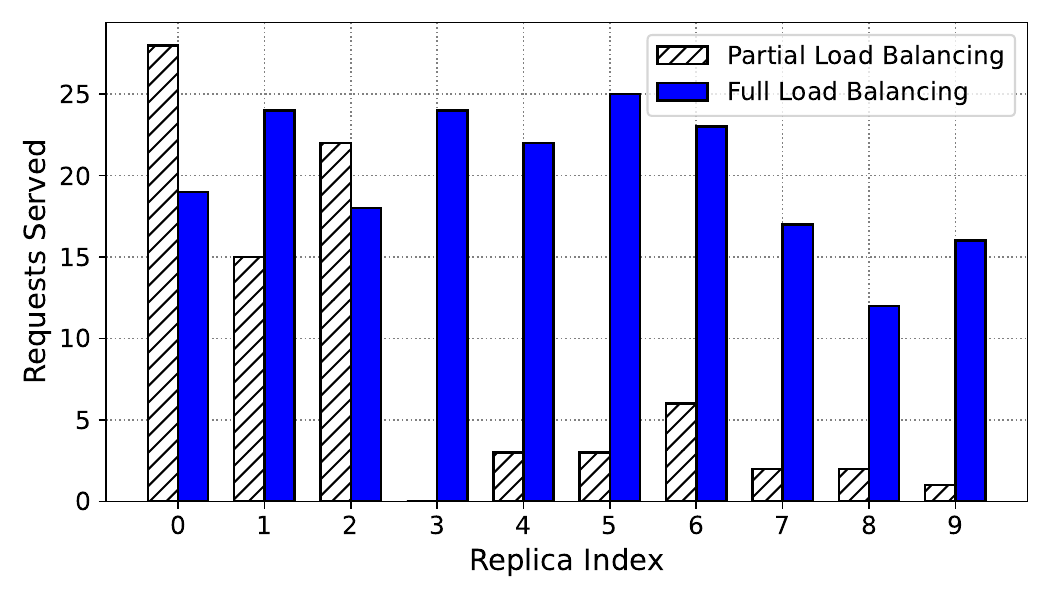}
         \vspace{-3mm}
		\caption{\label{fig:lb_results}Load balancing of QUIC traffic}
	\end{center}
       \vspace{-5mm}
\end{figure} 

To understand the impact of QUIC-aware load balancing in our system, we generate 200 client HTTP/3 requests using \textit{aioquic} library and compare two deployment scenarios: 
\begin{itemize}
    \item \textit{Case 1}: QUIC packets' deep analysis is not performed; hence, load balancing is only partial when some client requests target specific server IP addresses. In our test scenario, the majority of initial client packets are sent directly to IP addresses of Servers 1, 2, and 3, and the rest of the packets are sent to the VIP of the load balancer.
    \item \textit{Case 2}: QUIC packets are deep parsed and analyzed (as proposed in~\ref{sec:lbquic}). In this scenario, all initial QUIC client packets must be routed through the load balancer (i.e., the destination IP address must be the VIP address of the load balancer; otherwise, the packets are dropped).
\end{itemize}
In both scenarios, non-initial QUIC packets are allowed to bypass the load balancer, improving system performance while preserving session affinity to the server initially selected by the load balancer.

The results can be seen in Figure~\ref{fig:lb_results}. In the first scenario (Partial Load Balancing), the traffic is skewed at the first three servers. Uneven traffic distribution can degrade data center performance. Moreover, the load balancing bypass can be exploited by malicious users to intentionally disrupt cluster operation. 

In the second scenario (Full Load Balancing), deep parsing of QUIC packets allows the load balancer to identify initial QUIC packets with non-VIP destination addresses and drop them, preventing complete bypass of the load balancer. As a result, the distribution of traffic is significantly more even (in this implementation, we use ECMP hashing for selecting the server for every initial packet).

\subsection{0-RTT Attack Mitigation}
\begin{figure}
	\begin{center}
		\includegraphics[width=0.8\columnwidth]{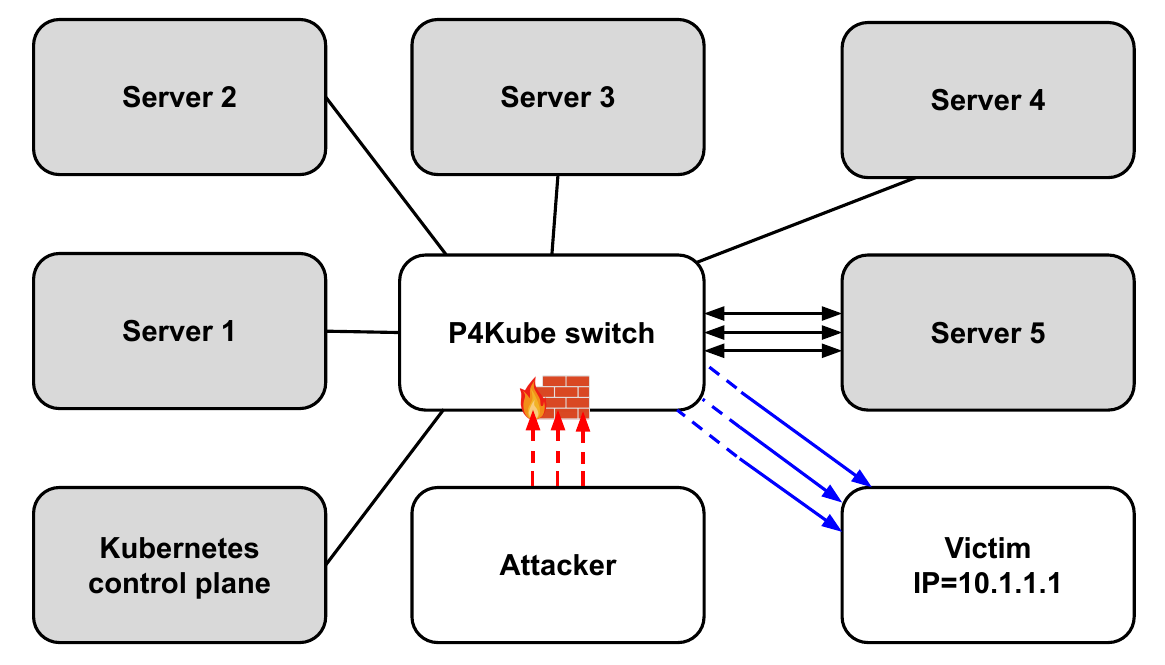}
		\caption{\label{fig:environment}Network topology of the experiment}
	\end{center}
\end{figure}

\begin{figure}[!t]

  \centering
  \subfloat[Attacker sends 0-RTT packets]{%
    \includegraphics[width=0.48\columnwidth]{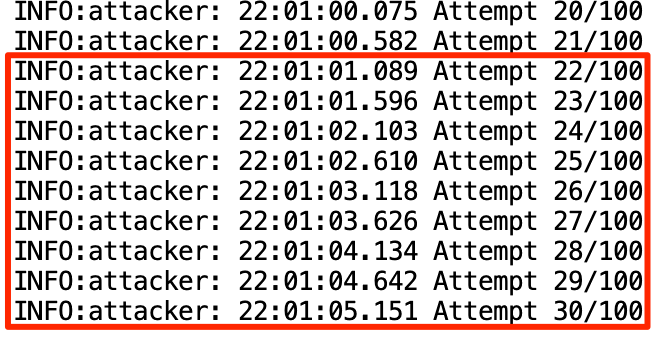}
  }\hfill
  \subfloat[Server (node5) stops receiving 0-RTT packets after \textit{22:01:00.801289}]{%
    \includegraphics[width=0.48\columnwidth]{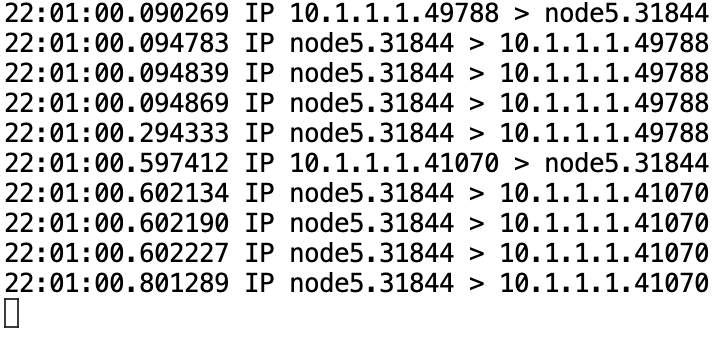}
  }
  \caption{Packet traces of the experiment. Packets highlighted in the red rectangle were blocked by the PISA switch}
  \label{fig:traces}
\end{figure}

We simulated the 0-RTT attack with IP spoofing and its mitigation in a Kubernetes-orchestrated cluster, deployed in an isolated lab testbed. The network topology of the experiment is shown in Figure~\ref{fig:environment}. To run the simulated attack, we used the implementation from~\cite{attack}, which follows the attack flow shown in Figure~\ref{fig:attack}. Specifically, the attacker first sends a legitimate request to obtain a TLS session ticket and then uses it to flood the server with 0-RTT packets with spoofed source IP addresses. In this experiment, PISA switch successfully blocks the attack, while forwarding legitimate traffic and 0-RTT packets before the attack is detected. To illustrate the experiment, we show the packet traces from the attacker's (spoofed IP$=$10.1.1.1) and server's terminals in Figure~\ref{fig:traces}.

\subsection{Deep QUIC Parsing in PISA Switch}
\begin{figure}
\vspace{-5mm}
	\begin{center}
		\includegraphics[width=0.5\columnwidth]{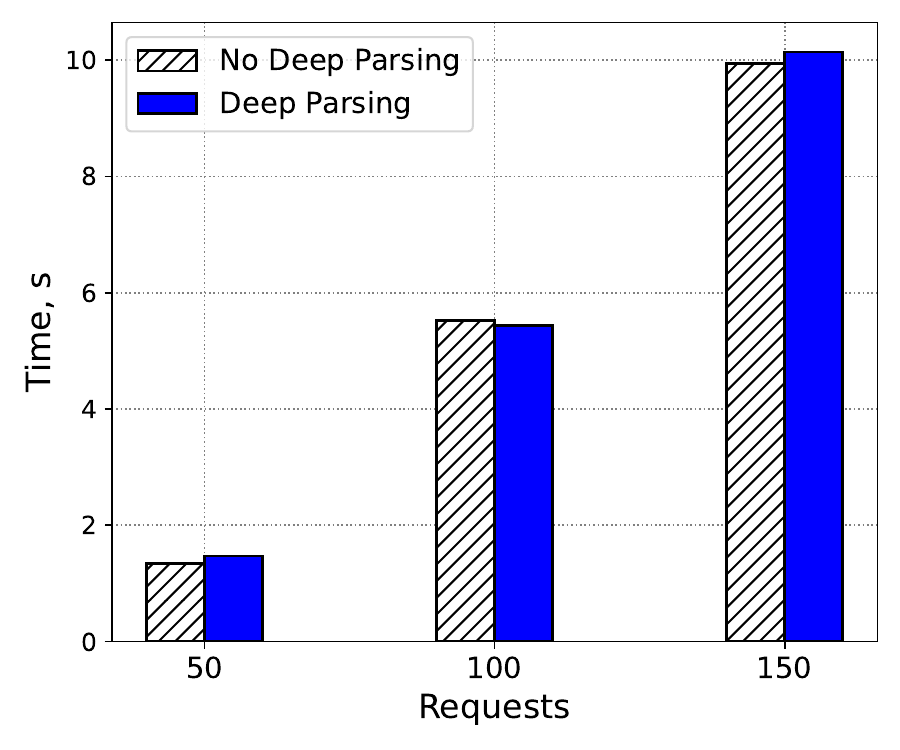}
        \vspace{-5mm}
		\caption{\label{fig:latency}Effect of deep QUIC parsing on HTTP/3 requests latency}
        
	\end{center}
\end{figure} 
The packet processing latency of the \textit{bmv2} switch is significantly higher than that of production-grade virtual or hardware PISA switches~\cite{chen2025enhancing}. Nevertheless, to obtain a preliminary estimate of the overhead of deep parsing of QUIC packets, including the processing of coalesced QUIC packets, we compared the round-trip time of HTTP/3 requests between (i) a load balancer that does not perform deep parsing for load balancer bypass prevention or 0-RTT attack mitigation, and (ii) the load balancer proposed in Section~\ref{sec:design}, which performs deep parsing. Using a topology similar to one shown in Figure~\ref{fig:environment}, we ran up to 150 parallel HTTP/3 requests from the client machine to the load balancer. As shown in Figure~\ref{fig:latency}, we found that deep parsing has little to no effect on the performance of QUIC flows, even in an emulated environment.



\section{Related work}
\label{sec:related}
\textit{SilkRoad}~\cite{miao2017silkroad} is a stateful load balancer that stores the mapping of active connections inside the match-action tables. On each new request, the data plane notifies the control plane software about the new flow and requires it to update the match-action tables. However, because packets arrive much faster than entries can be inserted into these tables, \textit{Silkroad} introduces transitional maps in data plane SRAM memory to temporarily handle connections that have not yet been registered by the control plane. 
\textit{CRAB}~\cite{kogias2020bypassing} supports stateless load-balancing by introducing a new TCP option, requiring modifications to the Linux kernel. Other stateless load balancers~\cite{pit2018stateless, barbette2020high, rizzi2021charon} achieve per-connection consistency by embedding the identifier of the mapped server into the least significant bits of the existing TCP timestamp field. However, TCP timestamp is an optional field and can be disabled by hosts' administrators. In addition, in the case of a large pool of servers (e.g., 1000), the modifications to the timestamp field can degrade the performance of TCP congestion control due to incorrect RTT measurements.

Using QUIC solves the problem of per-connection consistency thanks to the Connection ID (CID) present in the unencrypted fields of QUIC headers. A common approach is to embed the server identifier within a portion of the CID. The authors in~\cite{kistenmacher2025quic} explore several attacks on such a design, including the use of CIDs to track active connections, thus violating the QUIC privacy guarantees. 

In this work, we show how the QUIC flows can completely bypass the load balancer, with the exception of the first \textit{Initial} client packets. Our approach does not require changing CIDs. We also demonstrate how to mitigate 0-RTT IP spoofing attacks in this design. Unlike \textit{TrafficGrinder}~\cite{shahla2024trafficgrinder}, which limits 0-RTT ticket reuse via server-side modifications, our solution operates entirely in the network data plane by detecting and blocking 0-RTT spikes.
\section{Conclusion and Future Work}
\label{conclusion}
We design a programmable in-network load balancer for QUIC traffic. In our system, only the first packet of each flow needs to be handled by the load balancer algorithm; all subsequent packets can bypass it without affecting per-connection consistency. We demonstrate how to prevent complete load balancer bypass in this design and how to mitigate potential 0-RTT IP spoofing attacks. 
The implementation of the proposed system is available online\footnote{Link to the repository: https://github.com/gareging/P4Kube}. Future work may include testing this system on a physical hardware testbed and developing network-level mitigation strategies to address other QUIC vulnerabilities.

\bibliographystyle{IEEEtranS}
\bibliography{sample-base}
\end{document}